\documentclass{amsart}

\usepackage{amssymb}

\usepackage[english,francais]{babel}

\newtheorem{e-proposition}[theorem]{Proposition}

\newtheorem{e-definition}[theorem]{Definition\rm}

\newtheorem{theoreme}{Th\'eor\`eme}[section]

\newtheorem{proposition}[theoreme]{Proposition}

\def\og{\leavevmode\raise.3ex\hbox{$\scriptscriptstyle\langle\!\langle$~}}
\def\fg{\leavevmode\raise.3ex\hbox{~$\!\scriptscriptstyle\,\rangle\!\rangle$}}

\def \Z {\mathbb Z}
\def \R {\mathbb R}
\def \C {\mathbb C}

\DeclareMathOperator{\expect}{\mathbb{E}}
\DeclareMathOperator{\Sp}{Sp}
\DeclareMathOperator{\supp}{supp}

\title[Positivit\'e des exposants de Lyapounov]{Positivit\'e des exposants de Lyapounov pour un\\
op\'erateur de Schr\"odinger continu \`a valeurs matricielles}
\author{Hakim Boumaza}
\email{boumaza@math.jussieu.fr}
\address{Institut de Math\'ematiques de Jussieu, Universit\'e Paris 7 Denis Diderot, 2 place Jussieu, 75251 Paris, France}
\begin{document}
\maketitle

\begin{abstract}
\selectlanguage{francais}
Dans cette note, nous \'etudions un mod\`ele de type Anderson, continu et \`a valeurs matricielles.  Nous prouvons la stricte postivit\'e des deux plus grands exposants de Lyapounov associ\'es \`a ce mod\`ele, ainsi que leur s\'eparation, et ce pour toutes les \'energies dans $(2,+\infty)$ en dehors d'un ensemble discret de valeurs. Cela conduit \`a l'absence de spectre absolument continu dans $(2,+\infty)$. Les m\'ethodes, utilisant des r\'esultats de th\'eorie des groupes dus \`a Breuillard et Gelander, s'appliquent au cas singulier des distributions de Bernoulli. 
\vskip 0.5\baselineskip

\selectlanguage{english}
\noindent{\bf Abstract}
\vskip 0.5\baselineskip
\noindent {\bf Positivity of Lyapounov exponents for a continuous matrix-valued Schr\"odinger operator}

In this note, we study a continuous matrix-valued Anderson-type model. Both leading Lyapounov exponents of this model are proved to be positive and distincts for all energies in $(2,+\infty)$ except those in a discrete set, which leads to absence of absolutely continuous spectrum in $(2,+\infty)$. The methods, using group theory results by Breuillard and Gelander, allow for singular Bernoulli distributions. 
\end{abstract}

\selectlanguage{english}

\section*{Abridged English version}

Localization for Anderson models in dimension $d\geq 2$ is still an open problem if one looks for arbitrary disorder, especially for Bernoulli randomness. A possible approach to try to understand localization for $d=2$ is to discretize one direction. It leads to consider one-dimensional continuous Schr\"odinger operators, no longer scalar-valued, but now matrix-valued. 
What is already well understood is the case of dimension one scalar-valued continuous Schr\"odinger operators with arbitrary randomness including Bernoulli distributions (see \cite{stolz2}) and discrete matrix-valued Schr\"odinger operators also including the Bernoulli case (see \cite{goldsheid} and \cite{klein}). We aim at combining existing techniques for these cases to prove that for our model (\ref{model}), the Lyapunov exponents are all positive and distinct for all energies except those in a discrete set, at least if the energy is in $(2,+\infty)$ (see Theorem \ref{thmHA}). Due to Kotani's theory (see \cite{kotanis}) this result will imply the absence of absolutely continuous spectrum in the interval $(2,+\infty)$. To be applied, the results of \cite{kotanis} need to be combined with the suspension method described by Kirsch in \cite{kirsch}, as the operator we are studying here is not $\R$-ergodic but $\Z$-ergodic.
Our result is a modest first step before being able to prove the same kind of results for potentials of arbitrary dimension, not only $2\times 2$ matrices. But it is already interesting as, up to our knowledge, it is the first application of the work of Breuillard and Gelander on generating dense Lie subgroups in semisimple groups to a problem of separability of Lyapunov exponents. It also completes a first result of absence of absolutely continuous spectrum proved in \cite{stolzboumaza}. Indeed, in \cite{stolzboumaza} we only get positivity of Lyapunov exponents away from a countable set, which is enough to apply Kotani's theory but not enough to follow a multiscale analysis scheme to prove localization for our model (see \cite{stollmann}). The present result will allow this.
\vskip 2mm

In this Note, we begin by defining our model (\ref{model}), then we state precisely our result of positivity of Lyapunov exponents and absence of absolutely continuous spectrum for this model (Theorem \ref{thmHA}). Then we sketch the proof of this theorem. First we recall an algebraic criterion by Gol'dsheid and Margulis (see \cite{goldsheid}) which says that the Lyapunov exponents are all distinct and positive whenever the group $G_{\mu}(E)$ introduced at the section \ref{secmargulis} is Zariski-dense in the symplectic group $\Sp_2(\R)$. To prove Zariski-density we then follow a result of Breuillard and Gelander \cite{breuillarda} (Theorem \ref{breuillard}) on how to generate dense Lie subgroups of semisimple groups. To verify the assumptions of Theorem \ref{breuillard}, we first use simultaneous diophantine approximation to prove Proposition \ref{diop}. It is at this step that we have to restrict the values taken by the energy to the interval $(2,+\infty)$. Then we compute the logarithms of the elements of $\mathcal{O}$ (see Proposition \ref{breuillard}) given by Proposition \ref{diop} and we prove that they generate the whole Lie algebra $\mathfrak{sp}_{2}(\R)$, at least for energies not in a discrete set in $(2,+\infty)$. To do this we construct directly a family of $10$ linearly independent matrices in the Lie algebra generated by the logarithms. As $\mathfrak{sp}_{2}(\R)$ is of dimension $10$, this Lie algebra is in fact equal to $\mathfrak{sp}_{2}(\R)$ and we have satisfied the second assumption of Theorem \ref{breuillard}, which proves Theorem \ref{thmHA}.
\selectlanguage{francais}
\section{Introduction}\label{secintro}

La question de la localisation reste un probl\`eme ouvert pour les mod\`eles d'Anderson continus en dimension $d\geq 2$. Un moyen d'approcher de tels mod\`eles en dimension $2$ est de proc\'eder \`a  une discr\'etisation dans une direction, ce qui conduit \`a consid\'erer des mod\`eles d'Anderson continus \`a valeurs matricielles comme c'est le cas dans la pr\'esente note. Dans un premier article (voir \cite{stolzboumaza}), nous avons d\'ej\`a \'etudi\'e un tel mod\`ele: nous avons prouv\'e l'absence de spectre absolument continu pour des \'energies suffisamment grandes pour le mod\`ele (\ref{model}). Pour cela nous avons \'etabli la stricte positivit\'e des exposants de Lyapounov pour toute \'energie hors d'un ensemble d\'enombrable. Ce r\'esultat suffit pour prouver l'absence de spectre absolument continu, mais il ne permet pas ensuite d'\'etudier la r\'egularit\'e des exposants de Lyapounov et de la densit\'e d'\'etats int\'egr\'ee. Or cette  \'etape est essentielle en vue d'appliquer un sch\'ema d'analyse multi-\'echelle  pour prouver la localisation pour le mod\`ele (\ref{model}) (voir \cite{stollmann}). 

Pour \'etudier la r\'egularit\'e de la densit\'e d'\'etats int\'egr\'ee, il nous faut prouver que les exposants de Lyapounov associ\'es au mod\`ele (\ref{model}) sont deux \`a deux distincts et strictement positifs sur un intervalle, donc hors d'un ensemble \emph{discret} d'\'energies. Pour arriver \`a un tel r\'esultat, les techniques alg\'ebriques utilis\'ees dans \cite{stolzboumaza} se r\'ev\`elent insuffisantes. Nous les compl\'etons en utilisant un r\'esultat d\^u \`a Breuillard et
Gelander (voir \cite{breuillarda}) de densit\'e de sous-groupes de groupes de Lie semi-simples. 

L'int\'er\^et de ces techniques r\'eside aussi dans le fait qu'elles prennent en compte le cas limite des mod\`eles d'Anderson continus o\`u l'al\'ea appara\^{\i}t sous sa forme la plus faible, celle de variables al\'eatoires qui suivent une loi de Bernoulli. Jusque l\`a, seuls les cas d'op\'erateurs \`a valeurs scalaires (voir \cite{stolz2}) et d'op\'erateurs discrets \`a valeurs matricielles (voir \cite{goldsheid,klein}) avaient conduit \`a des preuves permettant de prendre en compte ce type d'al\'ea.


\section{Les r\'esultats}\label{secresults}

Le mod\`ele que nous \'etudions dans la pr\'esente Note est d\'efini par l'op\'erateur suivant: 
\begin{equation}\label{model}
H_{\omega}^A=-\frac{\mathrm{d}^2}{\mathrm{d}x^{2}} + 
\begin{pmatrix}
0 & 1 \\
1 & 0
\end{pmatrix}
+ \sum_{n\in\Z} 
\begin{pmatrix}
\omega_{1}^{(n)} \chi_{[0,1]}(x-n) & 0 \\
0 & \omega_{2}^{(n)} \chi_{[0,1]}(x-n)
\end{pmatrix}
\end{equation}
agissant dans $L^{2}(\R,\C^{2})$. Dans l'expression ci-dessus, $\chi_{[0,1]}$ est la fonction caract\'eristique de l'intervalle $[0,1]$, et $(\omega_{1}^{(n)})_{n\in\Z}$ et $(\omega_{2}^{(n)})_{n\in\Z}$ sont deux suites ind\'ependantes de variables al\'eatoires i.i.d.\ 
de distribution de probabilit\'e commune $\nu$ telle que $\{0,1\}\subset\supp\nu$.

Cet op\'erateur est une perturbation born\'ee de l'op\'erateur $(-\frac{\mathrm{d}^2}{\mathrm{d}x^{2}})
\oplus (-\frac{\mathrm{d}^2}{\mathrm{d}x^{2}})$. Il est donc autoadjoint dans l'espace de Sobolev $H^{2}(\R,\C^{2})$. \vskip 3mm 

Pour tout nombre r\'eel $E$, on associe au syst\`eme diff\'erentiel $H_{\omega}^{A}u=Eu$ la suite des matrices de transfert $(A_{n,2}^{\omega})_{n\in\Z}$ d\'efinie par les relations: 
\[
{^t}(u_{1}(n+1),u_{2}(n+1),u_{1}'(n+1),u_{2}'(n+1))=A_{n,2}^{\omega}(E).{^t}(u_{1}(n),u_{2}(n),u_{1}'(n),u_{2}'(n))
\]
o\`u $u=(u_{1},u_{2})$ est une solution du syst\`eme diff\'erentiel $H_{\omega}^{A}u=Eu$. Les matrices $A_{n,2}^{\omega}(E)$ appartiennent au groupe symplectique $\Sp_2(\R)$. Leurs exposants de Lyapounov, $\gamma_{1}(E),\ldots,\gamma_{4}(E)$, ont alors les propri\'et\'es de sym\'etrie suivantes: $\gamma_{1}=-\gamma_{4}$ et $\gamma_{2}=-\gamma_{3}$.

Pour plus de pr\'ecisions sur les exposants de Lyapounov dans le cas symplectique, on renvoie au chapitre $4$ de \cite{bougerol}. On est ainsi ramen\'e \`a n'\'etudier que les deux premiers exposants de Lyapounov.
Pour cet op\'erateur $H_{\omega}^{A}$, nous allons montrer le th\'eor\`eme suivant: 
\vskip 3mm

\begin{theoreme}\label{thmHA}
Il existe un ensemble discret $\mathcal{S}\subset \R$ tel que, pour tout $E\in (2,+\infty)$, $E\notin\mathcal{S}$, 
on ait
\[
\gamma_{1}(E)> \gamma_{2}(E) >0.
\]
En particulier, $H_{\omega}^{A}$ n'a pas de spectre absolument continu dans l'intervalle $(2,+\infty)$.
\end{theoreme}
\vskip 2mm

Ce th\'eor\`eme am\'eliore substantiellement le r\'esultat d\'ej\`a obtenu dans \cite{stolzboumaza} dans la mesure o\`u il nous permet d'obtenir l'existence d'intervalles ouverts sur lesquels la propri\'et\'e $\gamma_{1}(E)> \gamma_{2}(E) >0$ est vraie. Cela est essentiel pour pouvoir s'inspirer des techniques mises en place dans \cite{carmona}, puis d\'etaill\'ees dans \cite{stolz2} pour des op\'erateurs \`a valeurs scalaires et non matricielles. On pourra ainsi prouver un r\'esultat de r\'egularit\'e pour la densit\'e d'\'etats int\'egr\'ee associ\'ee \`a l'op\'erateur $H_{\omega}^{A}$. Un tel r\'esultat fera l'objet d'une publication ult\'erieure de l'auteur.

\section{Principe de la preuve}\label{secpreuve}

Nous pr\'esentons ici les principales id\'ees de la preuve du th\'eor\`eme \ref{thmHA}.

\subsection{Un crit\`ere alg\'ebrique de 
s\'eparation des exposants de Lyapounov}\label{secmargulis}

Tout d'abord on introduit pour tout $E$, \`a partir de la mesure $\nu$, une mesure $\mu_E$ sur le groupe symplectique $\Sp_2(\R)$. 
Pour $\Gamma$ bor\'elien de $\Sp_2(\R)$ on pose: $\mu_{E}(\Gamma)=\nu (\{ \omega \ |\ A_{0,2}^{\omega}(E) \in \Gamma \})$. Cette mesure $\mu_{E}$ est d\'efinie seulement \`a partir de $A_{0,2}^{\omega}(E)$ puisque les matrices $A_{n,2}^{\omega}(E)$ sont suppos\'ees i.i.d. On peut alors introduire le sous-groupe $G_{\mu}(E)$ de $\Sp_2(\R)$ engendr\'e par le support de la mesure $\mu_{E}$. Ce n'est autre que le sous-groupe engendr\'e par les quatre matrices $A_{0,2}^{(0,0)}(E), A_{0,2}^{(1,0)}(E), A_{0,2}^{(0,1)}(E),A_{0,2}^{(1,1)}(E)$ correspondant aux valeurs $0$ ou $1$ prises par $\omega_{1}^{(0)}$ et $\omega_{2}^{(0)}$.

Le crit\`ere essentiel de s\'eparation des exposants de Lyapounov est le suivant:
\vskip 3mm

\begin{theoreme}[Gol'dsheid et Margulis \cite{goldsheid}]\label{zariskidense}
Si le sous-groupe $G_{\mu}(E)$ est Zariski-dense dans $\Sp_2(\R)$, alors:
\(
\gamma_{1}(E)> \gamma_{2}(E) >0.
\)
\end{theoreme}
\vskip 2mm

Ce crit\`ere est utilis\'e dans \cite{stolzboumaza} o\`u nous n'\'etions parvenus \`a l'appliquer que pour un ensemble de valeurs de $E$ hors d'un ensemble d\'enombrable. 
Pour pallier cette insuffisance, il nous manquait le r\'esultat suivant qui permet de syst\'ematiser la preuve de Zariski-densit\'e de $G_{\mu}(E)$:
\vskip 3mm

\begin{theoreme}[Breuillard et Gelander \cite{breuillarda}]\label{breuillard}
Soit $G$ un groupe de Lie r\'eel connexe semi-simple d'al\-g\`ebre de Lie $\mathfrak{g}$. Il existe alors un voisinage de l'identit\'e $\mathcal{O} \subset G$, sur lequel $\log=\exp^{-1}$ est un diff\'eomor\-phisme bien d\'efini, et tel que $g_{1},\ldots g_{m}\in \mathcal{O}$ engendrent un sous-groupe dense de $G$ si et seulement si $\log(g_{1}),\ldots,\log(g_{m})$ engendrent $\mathfrak{g}$.
\end{theoreme}
\vskip 2mm

Ce crit\`ere nous donne le plan de la suite de la preuve. Tout d'abord nous allons v\'erifier que l'on peut construire \`a partir des matrices $A_{0,2}^{(0,0)}(E), A_{0,2}^{(1,0)}(E), A_{0,2}^{(0,1)}(E),A_{0,2}^{(1,1)}(E)$, quatre autres matrices qui sont dans $\mathcal{O}$. On calcule ensuite leurs logarithmes et on v\'erifie que ceux-ci engendrent l'alg\`ebre de Lie $\mathfrak{sp}_{2}(\R)$ du groupe $\Sp_2(\R)$.

\subsection{El\'ements proches de l'identit\'e dans $G_{\mu}(E)$}


\begin{proposition}\label{diop}
Soit $E\in (2,+\infty)$. Pour tout $\omega \in \{0,1\}^2$, il existe un entier $m_{\omega}(E)\geq 1$ tel que 
\[
(A_{0,2}^{\omega}(E))^{m_{\omega}(E)} \in \mathcal{O}.
\]
\end{proposition}

On commence par pr\'eciser l'expression des matrices de transfert. On pose: 
\[
M_{\omega^{(0)}}=
\begin{pmatrix}
\omega_{1}^{(0)} & 1 \\
1 & \omega_{2}^{(0)}
\end{pmatrix}= 
S_{\omega^{(0)}}\begin{pmatrix}
\lambda_{1}^{\omega^{(0)}} & 0 \\
0 & \lambda_{2}^{\omega^{(0)}}
\end{pmatrix}S_{\omega^{(0)}}^{-1},
\]
o\`u $S_{\omega^{(0)}}$ est orthogonale et o\`u les nombres r\'eels $\lambda_{1}^{\omega^{(0)}}\leq \lambda_{2}^{\omega^{(0)}}$ sont les valeurs propres de $M_{\omega^{(0)}}$. On peut calculer explicitement toutes ces quantit\'es. Si $E>2$, alors $E$ est plus grand que toute valeur propre de $M_{\omega^{(0)}}$, et on obtient l'expression suivante, o\`u l'on note $r_i=r_i(E,\omega^{(0)}):= \sqrt{E-\lambda_i^{\omega^{(0)}}}$ pour $i=1,2$: 
{\footnotesize \begin{equation} \label{expltransfer}
A_{0,2}^{\omega^{(0)}}(E)=
\begin{pmatrix}
S_{\omega^{(0)}} & 0 \\
0 & S_{\omega^{(0)}}
\end{pmatrix}
\begin{pmatrix}
\cos r_1 & 0 & \tfrac{1}{r_{1}} \sin r_1 & 0 \\[1mm]
0 & \cos r_2 & 0 & \tfrac{1}{r_{2}} \sin r_2 \\[1mm]
-r_{1} \sin r_1 & 0 & \cos r_1 & 0 \\[1mm]
0 & -r_{2} \sin r_2 & 0 & \cos r_2
\end{pmatrix}
\begin{pmatrix}
S_{\omega^{(0)}}^{-1} & 0 \\
0 & S_{\omega^{(0)}}^{-1}
\end{pmatrix}
\end{equation} }
On fixe $E\in(2,+\infty)$ et $\omega=(\omega_1,\omega_2)\in \{ 0,1\}^{2}$. Soit $M$ un nombre r\'eel $>1$. Par approximation diophantienne simultan\'ee (voir \cite{schmidt}), on voit qu'il existe $m_{\omega}(E) \in\Z$, $1\leq m_{\omega}(E) \leq M$ et $(x_{1}, x_{2}) \in\Z^{2}$ tels que: 
\[
|r_{1}m_{\omega}(E)-2x_{1}\pi|<2\pi M^{-\frac{1}{2}}\ \text{ et }\ |r_{2}m_{\omega}(E)-2x_{2}\pi|<2\pi M^{-\frac{1}{2}}.
\]
En prenant alors $M$ assez grand  pour que la boule de centre l'identit\'e et de rayon $2\pi M^{-\frac{1}{2}}$ soit contenue dans $\mathcal{O}$, on en d\'eduit que $(A_{0,2}^{\omega}(E))^{m_{\omega}(E)} \in \mathcal{O}$. On a obtenu les \'el\'ements voulus de $G_{\mu}(E)$ qui sont dans $\mathcal{O}$. Il est tr\`es important pour la suite de pr\'eciser que l'on peut fixer $M$ ind\'ependant de $E$ et de $\omega$. \qed

\subsection{L'alg\`ebre de Lie $\mathcal{A}$}

\begin{proposition}\label{logen}
Les logarithmes des matrices $(A_{0,2}^{\omega}(E))^{m_{\omega}(E)}$ engendrent 
l'alg\`ebre de Lie $\mathfrak{sp}_{2}(\R)$ pour tous les $E\in(2,+\infty)$ hors d'un ensemble discret $\mathcal{S}$.
\end{proposition}
\vskip 2mm

Tout d'abord, on a l'expression suivante pour $LA^{\omega}:=\log((A_{0,2}^{\omega}(E))^{m_{\omega}(E)})$: 
{\scriptsize 
\begin{equation}
LA^{\omega}=
\begin{pmatrix}
S_{\omega^{(0)}} & 0 \\
0 & S_{\omega^{(0)}}
\end{pmatrix}\!
\begin{pmatrix}
0 & 0 & m_{\omega}(E)-\frac{2\pi x_{1}}{r_{1}} & 0 \\
0 & 0 & 0 & m_{\omega}(E)-\frac{2\pi x_{2}}{r_{2}} \\
m_{\omega}(E)r_{1}^{2}+2\pi r_{1} x_{1} & 0 & 0 & 0 \\
0 & m_{\omega}(E)r_{2}^{2}+2\pi r_{2} x_{2} & 0 & 0 
\end{pmatrix}\!
\begin{pmatrix}
S_{\omega^{(0)}}^{-1} & 0 \\
0 & S_{\omega^{(0)}}^{-1}
\end{pmatrix}
\end{equation}}%
o\`u les $x_i=x_i^{\omega}(E):=\frac{1}{2} \expect\bigl(\frac{m_{\omega}(E)r_{i}}{\pi} +\frac{1}{2}\bigr)$, pour $i=1,2$, sont comme dans la preuve de la proposition \ref{diop}.
On note $\mathcal{A}\subset\mathfrak{sp}_{2}(\R)$ la sous-alg\`ebre de Lie engendr\'ee par les quatre logarithmes $LA^{\omega}$. On commence par prouver que les $4$ crochets de Lie $[LA^{(1,0)},LA^{(0,0)}]$, $[LA^{(0,1)},LA^{(0,0)}]$, $[LA^{(1,0)},LA^{(1,1)}]$ et $[LA^{(0,1)},LA^{(1,1)}]$  forment une famille libre dans l'espace
$V_1$ de dimension $4$ suivant:
\[
V_{1}:=\left\{Z=\bigl(\begin{smallmatrix}
A & 0 \\
0 & -{}^tA
\end{smallmatrix}\bigr)\bigm\vert 
A=\bigl(\begin{smallmatrix}
a & b\\
c & d
\end{smallmatrix}\bigr)\in\mathcal{M}_{2}(\R)\right\} \subset \mathfrak{sp}_{2}(\R),
\]
et ce pour tout $E$ hors d'un ensemble discret $\mathcal{S}_{1}$. En effet, on peut calculer le d\'eterminant form\'e des $4$ vecteurs colonnes repr\'esentant les coefficients des blocs sup\'erieurs gauches $2\times 2$ des $4$ crochets de Lie. Ce d\'eterminant est une fonction $d(E,m_{\omega}(E),x_i^{\omega}(E))$. Elle n'est pas directement analytique car les $m_{\omega}(E)$ et les $x_i^{\omega}(E)$ ne le sont pas a priori. Mais, si l'on fixe un intervalle born\'e $I$ (ouvert ou ferm\'e) contenu dans $(2,+\infty)$, et un nombre $M$
comme dans la preuve de la proposition \ref{diop}, les entiers $m_{\omega}(E)$ et $x_i^{\omega}(E)$ ne prennent qu'un nombre fini de valeurs lorsque $E$ parcourt $I$. Or, si on fixe ces entiers, on obtient bien une fonction analytique de $E$. Ainsi, l'ensemble
\[
\mathcal{S}_I=\{ E \in I\mid d(E,m_{\omega}(E),x_i^{\omega}(E))=0 \}=\bigcup_{p,k_{i}} \{ E \in I\mid d(E,p,k_{i})=0 \}
\]
est discret comme r\'eunion finie d'ensembles discrets. Comme cela vaut pour tout sous-intervalle born\'e de
$(2,+\infty)$, on en d\'eduit que l'ensemble $\mathcal{S}_{1}$ des $E\in (2,+\infty)$ tels que 
$d(E,m_{\omega}(E),x_i^{\omega}(E))=0$ est discret.

Si on fixe $E\in (2,+\infty)$, $E\notin\mathcal{S}_{1}$, on peut utiliser le fait qu'alors $V_{1} \subset \mathcal{A}$ pour dire que les matrices $Z_{1}$, $Z_{2}$ et $Z_{3}$ correspondant \`a 
$(a,b,c,d)=(1,0,0,0)$, $(0,0,0,1)$ et $(1,1,1,1)$ sont aussi dans $\mathcal{A}$. On prouve alors, comme pr\'{e}c\'{e}demment, que $LA^{(1,0)}-LA^{(0,0)}$, $LA^{(1,0)}-LA^{(1,1)}$, $LA^{(0,1)}-LA^{(0,0)}$, $[LA^{(1,0)}-LA^{(0,0)},Z_{1}]$, $[LA^{(1,0)}-LA^{(1,1)},Z_{2}]$ et $[LA^{(0,1)}-LA^{(0,0)},Z_{3}]$ forment une famille libre de $6$ \'el\'ements dans l'espace $V_2$ de dimension $6$ suivant: 
\[
V_{2}:=\left\{\bigl(\begin{smallmatrix}
0 & C \\
B & 0
\end{smallmatrix}\bigr)\bigm\vert B,C\in \mathcal{M}_{2}(\R)\ \mathrm{sym\acute{e}triques} \right\} \subset \mathfrak{sp}_{2}(\R),
\]
et ce pour tout $E$ hors d'un ensemble discret $\mathcal{S}_{2}$. 

On pose $\mathcal{S}=\mathcal{S}_{1} \cup \mathcal{S}_{2}$. 
Comme $\mathfrak{sp}_{2}(\R)=V_{1} \oplus V_{2}$, on vient donc d'exhiber, pour tout $E\in (2,+\infty)\setminus \mathcal{S}$, une famille libre de $10$ matrices dans $\mathcal{A}$. Comme $\dim\mathfrak{sp}_{2}(\R)=10$ et que $\mathcal{A}$ en est un sous-espace, cela prouve l'\'egalit\'e $\mathcal{A}=\mathfrak{sp}_{2}(\R)$ pour ces valeurs de $E$. Ainsi, pour tout $E\in(2,+\infty)$, $E\notin\mathcal{S}$ les matrices $LA^{\omega}$ engendrent l'alg\`ebre de Lie $\mathfrak{sp}_{2}(\R)$.\hfill\qed
\vskip 3mm

\subsection{Fin de la preuve du th\'eor\`eme \ref{thmHA}}\label{secendproof}

On peut appliquer le th\'eor\`eme \ref{breuillard} pour $E\in (2,+\infty) \setminus \mathcal{S}$. Les matrices $((A_{0,2}^{\omega}(E))^{m_{\omega}(E)})$ engendrent donc un sous-groupe dense dans $\Sp_2(\R)$. Comme ce sous-groupe est contenu dans $G_{\mu}(E)$, on en d\'eduit que $G_{\mu}(E)$ est dense dans $\Sp_2(\R)$. En particulier il est Zariski-dense. Le th\'eor\`eme \ref{zariskidense} s'applique donc, ce qui prouve la s\'eparation et la stricte positivit\'e des exposants de Lyapounov. 

Pour l'assertion sur l'absence de spectre absolument continu, on se r\'ef\`ere \`a la th\'eorie de Kotani
pour les op\'erateurs $\R$-ergodiques (\cite{kotanis}, th\'eor\`eme $7.2$). Comme ici, $H_{\omega}^{A}$ est $\Z$-ergodique, et non $\R$-ergodique, on fait appel aux techniques de suspension de Kirsch dans \cite{kirsch} pour appliquer la th\'eorie de Kotani aux op\'erateurs $\Z$-ergodiques.

\section*{Remerciements}

L'auteur tient \`a remercier vivement Anne Boutet de Monvel et G\"unter Stolz pour leurs nombreux conseils 
et suggestions, et pour toute l'attention qu'ils ont port\'ee \`a son travail. 


\end{document}